\documentclass[preprint,12pt,nofootinbib]{revtex4}
\usepackage{amsmath, amssymb, amsfonts, latexsym, epsfig}
\usepackage{array}
\usepackage{epstopdf}
\usepackage{bm}
\usepackage{cancel}
\usepackage{color}
\usepackage{epsfig}
\usepackage{float}
\usepackage{graphicx}
\usepackage{mathrsfs}
\usepackage{slashed}
\usepackage{subfig}
\usepackage{textcomp}
\usepackage{times}
\usepackage{url}
\usepackage{graphicx}
\usepackage{cancel}
\usepackage{textcomp}
\usepackage{amsmath, amssymb, amsfonts, latexsym, epsfig}
\usepackage{mathrsfs}

\def\beq{\begin{equation}}
\def\eeq{\end{equation}}
\def\bea{\begin{eqnarray}}
\def\eea{\end{eqnarray}}

\begin{document}
\title{Light Higgsino from $A_t$ Dilemma in Rare $B$-decays}
\author{Ming-xing Luo}
\email{mingxingluo@zju.edu.cn}
\author{Kai Wang}
\email{wangkai1@zju.edu.cn}
\author{Tao Xu}
\email{taoxu@zju.edu.cn}
\author{Liangliang Zhang}
\email{zhngliang87@zju.edu.cn}
\author{Guohuai Zhu}
\email{zhugh@zju.edu.cn}
\affiliation{Zhejiang Institute of Modern Physics, Department of Physics, Zhejiang University, Hangzhou, Zhejiang 310027, China}

\begin{abstract}
In the Minimal Supersymmetric Standard Model (MSSM), large chiral symmetry breaking term $A_t$, which plays an important role in Higgs mass, may significantly contribute in flavor changing neutral current (FCNC) processes $B\to X_{s}\gamma$ and $B_{s}\to \mu^{+}\mu^{-}$. Though the above processes can both be categorized as $b\rightarrow s$ transitions, the two rare decays behave completely different in MSSM. With an on-shell photon in the final state, helicity of initial state $b$-quark and final state $s$-quark must be flipped in
 $B\to X_{s}\gamma$, which corresponds to the simultaneous breaking of chiral symmetry and electroweak symmetry. The common feature is shared by fermion mass generation. Same as radiative mass generation in MSSM, Peccei-Quinn and $R$ symmetry breaking contributions, for example from a Higgsino-stop loop when $\mu A_t<0$, may significantly cancel the contribution from charged Higgs and reduce the prediction of  $B\to X_{s}\gamma$. For the latter process, including Babu-Kolda FCNC proportional to $\mu A_t$, $B_{s}\to \mu^{+}\mu^{-}$ is mediated by a scalar $H_d$ boson which corresponds to chiral symmetry breaking. In addition, as a result of interference among the Higgs extension sector and $Z$ contributions, in the region of $\mu A_t <0$ which is favored by $B\to X_{s}\gamma$, there may simultaneously exist large enhancement in $B_{s}\to \mu^{+}\mu^{-}$. However, we  still find viable parameter region with light Higgsino of a few hundreds GeV when charged Higgs contribution is not negligible with $M_A$ of 400~GeV.

\end{abstract}
\maketitle
\section{Introduction}

A standard model(SM)-like Higgs boson with a mass of approximately 125~~{\rm GeV} has been discovered by both ATLAS and CMS collaborations at CERN Large Hadron Collider (LHC) \cite{Higgs_discover,Higgs_discover1}. The discovery of this seemingly last piece of SM has significantly improved our understanding of the mechanism of spontaneously electroweak symmetry breaking. On the other hand, neither the mass of the Higgs boson nor the driving force of electroweak symmetry breaking is explained within the SM.
The above questions in addition to the quadratic divergence in quantum correction to Higgs boson mass
has been driving the studies of beyond SM physics in the last three decades and the direct searches of these proposed models is one of leading tasks of LHC.
Supersymmetry is one of the most elegant solutions to the above questions. The Higgs boson mass is protected by the supersymmetry from quadratic divergence correction and the electroweak symmetry breaking is driven by the radiative generated potential as Coleman-Weinberg mechanism \cite{Coleman:1973jx}.  The leading SM contribution to Higgs quartic coupling is from top quark due to large top quark Yukawa coupling but the sign is opposite to the required contribution \cite{Dermisek:2013pta}. Introduction of the supersymmetric partners resolves
the problem. Besides the solutions to EWSB, minimal supersymmetric standard model (MSSM) also predicts unification of gauge couplings at high scale and provides natural dark matter candidate at the same time.

MSSM is naturally a type-II two-Higgs-Doublet model (2HDM). Holomorphic condition forbids the $\bar{H}$ in the superpotential. Cancellation of the mixed $[SU(2)]^{2}U(1)_{Y}$ gauge anomaly as well as the Witten anomaly also requires a pair of vector-like Higgsinos. The MSSM superpotential is then
\beq
{\cal W}=y_{u}Qu^{c}H_{u}+y_{d}Qd^{c}H_{d}+y_{e}\ell e^{c} H_{d} + \mu H_{u} H_{d}~.
\label{mssm}
\eeq
Therefore supersymmetry searches at the colliders consist of both the indirect search of extended Higgs sector and the direct search of supersymmetric partners.
The search of Higgs extension completely relies on mass spectrum and couplings of $H$, $A$ and $H^{\pm}$.
Lower $M_{A}$ and larger $\tan\beta$ will significantly enhance the discovery potential of extra Higgs bosons  at the LHC.

Flavor physics plays important roles in testing BSM physics  due to its sensitivity to interference of loop effects. Type-II 2HDM receives stringent constraints from rare
decay processes, in particular the $B\to \tau\nu_{\tau}$, $B\to X_{s}\gamma$ and $B_{s}\to \mu^{+}\mu^{-}$.
However, in MSSM, contributions from supersymmetric partners may interfere with
extended Higgs bosons' contribution except the $B\to \tau\nu_{\tau}$ which is dominated
by tree-level interference between SM and the new $H^{\pm}$ state. Consequently, taking into account the destructive interference from sparticles, parameter region of lower $M_{A}$ and larger $\tan\beta$ which is excluded in Type-II 2HDM may still be viable under the stringent bounds of $B\to X_{s}\gamma$ and $B_{s}\to \mu^{+}\mu^{-}$.

The two $b\to s$ transitions, $B\to X_{s}\gamma$ and $B_{s}\to \mu^{+}\mu^{-}$, also behave  differently in MSSM. $B\to X_{s}\gamma$ only involves a $b\to s$ transition which arises from
the $W$-loop in SM. In Type-II 2HDM, the introduction of charged Higgs states enhances the
amplitude due to a constructive interference between the $H^{\pm}$-loop and $W^{\pm}$-loop.
However, besides the $b\to s$ transition, pseudo-scalar decay $B_{s}\to \mu^{+}\mu^{-}$ is mediated by an off-shell $Z$ boson. The neutral Higgs boson $H_{d}$-mediated processes destructively interfere with such SM contribution and make the $B_{s}\to \mu^{+}\mu^{-}$ smaller than the SM prediction. In MSSM, the leading supersymmetric contribution to $B_{s}\to \mu^{+}\mu^{-}$ is also mediated by the $H_{d}$-like neutral Higgs boson. Therefore, when including the supersymmetric contribution to $b\to s$ transition, if the new contribution is destructively
interfering with charged Higgs diagram in $B\to X_{s}\gamma$, the total prediction to  $B_{s}\to \mu^{+}\mu^{-}$ may even be  enhanced.

The helicity between initial $b$-quark state and final $s$-quark state in $b\to s \gamma$
must be flipped due to the spin-1 photon in final state. Helicity flip is a consequence of simultaneous breaking of chiral symmetry and electroweak gauge symmetry. In SM, such contribution appears as a $m_{b}$ mass insertion in the amplitude.
The same feature is shared by the radiative correction to the fermion mass generation. Therefore, one would expect a correlation between $B\to X_{s}\gamma$ and correction in $b$-quark mass generation \cite{Wang:2013pxa}.

In Eq.\ref{mssm}, $b$-quark mass arises from Yukawa coupling
$y_{b} Q d^{c} H_{d}$ and {\it vev} of $H_{d}$ Higgs.
On the other hand,  gauge invariant $Q d^{c} \bar{H}_{u}$ coupling could exist in  Lagrangian to give loop correction\cite{Hall:1993gn}. Hence, $b$-quark mass generation is from both $v_{d}$ and $v_{u}$ as  $m_{b}= y_{b} v_{d}+\Delta m_{b} (v_{u})$~. Then the yukawa coupling can be defined as $y_{b}=\frac{m_{b}\tan\beta}{v\left(1+\epsilon\tan\beta\right)}$
with $\tan\beta=\frac{v_{u}}{v_{d}}$\cite{Djouadi:2005gj}. This $y_{b}$ coupling describes the coupling
of bottom quark to the down-type neutral Higgs $H_{d}$ in the interaction
basis. $\epsilon$ stands for the correction from the up-type neutral
Higgs with an effective vertex $Qd^{c}\bar{H}_{u}$.
Existence of such supersymmetric mass correction is a leading source of flavor violation in
MSSM since one cannot diagonalize the masses of the quarks in the same basis as their Yukawa couplings~\cite{Babu:1999hn,Kane:2003wg}.

A powerful tool to categorize such supersymmetric corrections is through symmetry approach.
The term $Q d^{c} \bar{H}_{u}$ breaks two global $U(1)$ symmetries, $R$ symmetry and Peccei-Quinn (PQ) symmetry. A global $U(1)$ $R$-transformation is defined as a rotation over the anti-commuting coordinates (Grassmann variables) $\theta$ and $\bar{\theta}$.
$R$-symmetry is the chiral symmetry protecting the Majorana gaugino mass and is broken when gaugino mass is generated along the supersymmetry breaking~\cite{Nelson:1993nf}.
Type-II 2HDM intrinsically contains a global PQ symmetry \cite{Dine:1981rt,Zhitnitsky:1980tq}.  If the bare $\mu$-term in Eq.\ref{mssm} is forbidden by a $U(1)_{X}$ under which $H_{u}H_{d}$ is not invariant, such $U(1)_{X}$ can then be identified as a PQ symmetry with non-vanishing mixed QCD-$U(1)_{X}$ anomaly $
A_{[SU(3)_{C}]^{2}U(1)_{X}} ={N_{\rm G}} (2 q+ u+d)/2 = -{N_{\rm G}}(h_{u}+h_{d})/2$~.
The $\mu$-term which corresponds to the Higgsino mass term explicitly breaks the PQ symmetry.
We give corresponding PQ and $R$ charge assignments under convention consistent with $SU(5)$ in Table \ref{ccc}.
\begin{table}[h]
\begin{center}
\caption{Charge assignment under $R$-symmetry and Peccei-Quinn symmetry.}
\begin{tabular*}{1.0\textwidth}{@{\extracolsep{\fill}} c || c c c  c c c c | c | c  }
\hline
 & $Q$ & $u^{c}$ & $e^{c}$ & $d^{c}$ & $\ell$ & $H_{u}$ & $H_{d}$ & $\theta$  & $Q d^{c} \bar{H}_{u}$\\
\hline\hline
$R$-charge &  ${1/ 5}$ & ${1/ 5}$ & ${1/ 5}$ & $7/ 5$ & ${7/ 5}$ & $8/ 5$ & $2/ 5$ & 1 & 0\\
PQ   & 0 & 0 & 0 & $-$1 & $-$1 & 0 & 1 & 0 &  $-$1 \\
\hline
\end{tabular*}
\label{ccc}
\end{center}
\end{table}
Clearly $Q d^{c} \bar{H}_{u}$ term breaks the PQ symmetry. Taking two fermionic components of $Q$ and $u^{c}$ field,  the $R$-invariant condition in Lagrangian is of $R$-charge 2 while the term $Q d^{c} \bar{H}_{u}$ is 0, so it breaks $R$-symmetry. For instance, a typical correction as $Q d^{c}\bar{H}_{u}$ from Higgsino-stop loop is proportional to $\mu A_{t}$ in which $\mu$ breaks the PQ symmetry while $A_{t}$ breaks the chiral symmetry $U(3)_{Q}\times U(3)_{u}$ as well as $R$-symmetry.

MSSM imposes strong constraints on the Higgs mass spectrum. When $M_{A}> M_{Z}$, at tree level, the lighter CP-even Higgs boson mass is below $m_{Z}$ as $m_{Z} \cos{2\beta}$~.
However, the Higgs boson mass also receives large radiative corrections from strong chiral symmetry $U(3)_{Q}\times U(3)_{u}$ breaking sources, such as the Higgs couplings to the top quarks, top Yukawa coupling $y_{t}$ and to their spin-0 SUSY partners, the trilinear coupling $A_{t}$. For instance, the 1-loop precise correction is \cite{Carena:1995bx} :
\begin{equation}
\Delta m_{h}^{2}\simeq \frac{3m_{t}^{4}}{4\pi^{2}v^{2}}\left[\log\frac{M_{\mathrm{SUSY}}^{2}}{m_{t}^{2}}+\frac{\tilde{A}_{t}^{2}}{M_{\mathrm{SUSY}}^{2}}\left(1-\frac{\tilde{A}_{t}^{2}}{12M_{\mathrm{SUSY}}^{2}}\right)\right],\label{eq:Higgs mass}
\end{equation}
where $M_{\mathrm{SUSY}}^{2}=m_{\tilde{t}_{1}}m_{\tilde{t}_{2}}$
is the averaged stop mass square and $\tilde{A}_{t}=A_{t}-\mu\cot\beta$.

Large chiral symmetry breaking which gives rise to the Higgs boson mass at the same time may also contribute significantly to the flavor physics as well as the Yukawa couplings.
Such correlation gives rich phenomenology in MSSM and is the focus of our paper.
We discuss in detail the flavor physics in this scenario in the next section.
In the above discussion, the third generation squark stop always appear in the loop.
As a result of the large yukawa coupling and trilinear coupling, running mass of
stop is typically lighter than the other squarks. In the third section, we focus on
the Higgsino-stop loop contribution and study the numerical result of viable parameter space. We then conclude in the final section.

\section{$b\rightarrow s$ transition and B-meson decays in MSSM}
Rare B-meson decays such as $B\rightarrow X_{s}\gamma$ and $B{}_{s}\rightarrow\mu^{+}\mu^{-}$
include a $b\rightarrow s$ transition, which is loop suppressed with additional CKM suppression in the SM. Experimentally these decays have been observed. For $B{}_{s}\rightarrow\mu^{+}\mu^{-}$ decay, a combined result of LHCb and CMS gives\cite{CMS:2014xfa}
\beq
\mathcal{B}\left(B{}_{s}\rightarrow\mu^{+}\mu^{-}\right)=\left( 3.1\pm0.7\right)\times10^{-9}~.
\label{expbmumu}
\eeq
$B\rightarrow X_{s}\gamma$ decay has been measured precisely for a photon energy cut of $E_\gamma>1.6$ GeV in the B-meson rest frame\cite{Lees:2012ufa,Limosani:2009qg,Chen:2001fja}. The current experimental world average reads as\cite{Amhis:2014hma}
\beq
\mathcal{B}\left(B\rightarrow X_{s}\gamma\right)=\left(3.43\pm0.21\pm0.07\right)\times10^{-4}~.
\label{expbsgamma}
\eeq
These measurements are in good agreement with the
SM predictions \cite{Bobeth:2013uxa,Misiak:2006zs}
\beq
\mathcal{B}\left(B{}_{s}\rightarrow\mu^{+}\mu^{-}\right)_{SM}=\left(3.65\pm0.23\right)\times10^{-9}~,
\label{smbmumu}
\eeq
\beq
\mathcal{B}\left(B\rightarrow X_{s}\gamma\right)_{SM}=\left(3.15\pm0.23\right)\times10^{-4}~.
\label{smbsgamma}
\eeq
This indicates that there's little room for new physics contributions
beyond the SM in $b \to s$ transition. But this does not exclude completely the possibility of light sparticles, as we will show later. In addition,  the pure leptonic decay $B\rightarrow\tau\nu_\tau$ also set up strict bound on BSM physics\cite{Adachi:2012mm,Lees:2012ju}. However, this decay channel is sensitive only on
 the extended Higgs sector, we prefer to apply $B\rightarrow\tau\nu_\tau$ bound to the numerical scan in the last section. In this section, we shall discuss in detail $B\rightarrow X_{s}\gamma$ and $B{}_{s}\rightarrow\mu^{+}\mu^{-}$ decays and their implications in MSSM.

Let's first discuss the $b\rightarrow s$ transition in Type II 2HDM.
In this circumstance, the $B\rightarrow X_{s}\gamma$ decay is enhanced by a $H^{+}$-$t$ loop
which depends only on $M_A$ and $\tan\beta$. So the
charged Higgs should be relatively heavy to avoid violating the experimental bound of $\mathcal{B}\left(B\rightarrow X_{s}\gamma\right)$. But for $B_{s}\rightarrow\mu^{+}\mu^{-}$ decay, the 2HDM
contributions interfere destructively with the SM \cite{Hewett:1988mi,He:1988tf,Logan:2000iv}. In case of large $\tan\beta$, the charged Higgs contribution could be much larger than the SM one. Therefore varying $\tan\beta$ from small to large, the branching ratio will first decrease to a minimum about half of the SM prediction and then increase monotonically.

\begin{figure}[ht]
\centering
\subfloat[$B\rightarrow X_{s}\gamma$]{\includegraphics[scale=0.8]{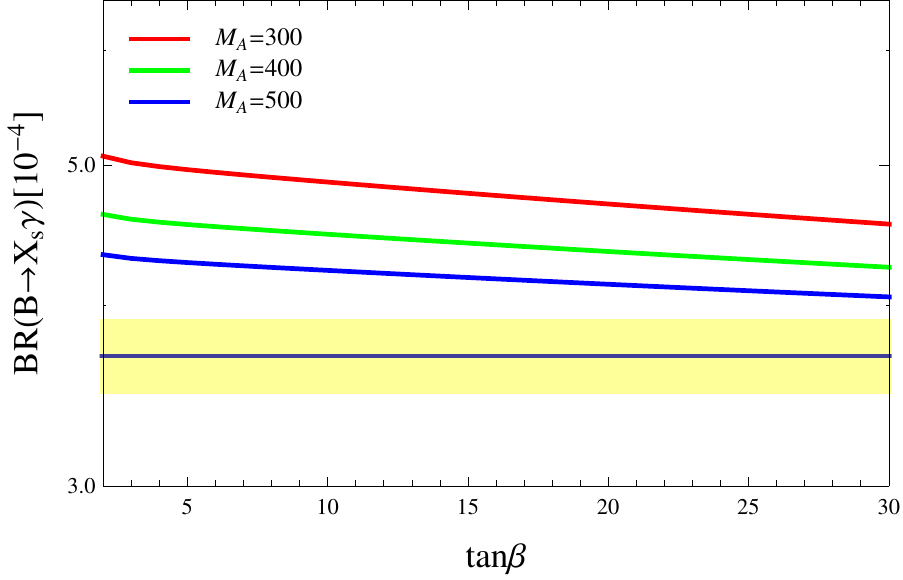}}\qquad
\subfloat[$B{}_{s}\rightarrow\mu^{+}\mu^{-}$]{\includegraphics[scale=0.8]{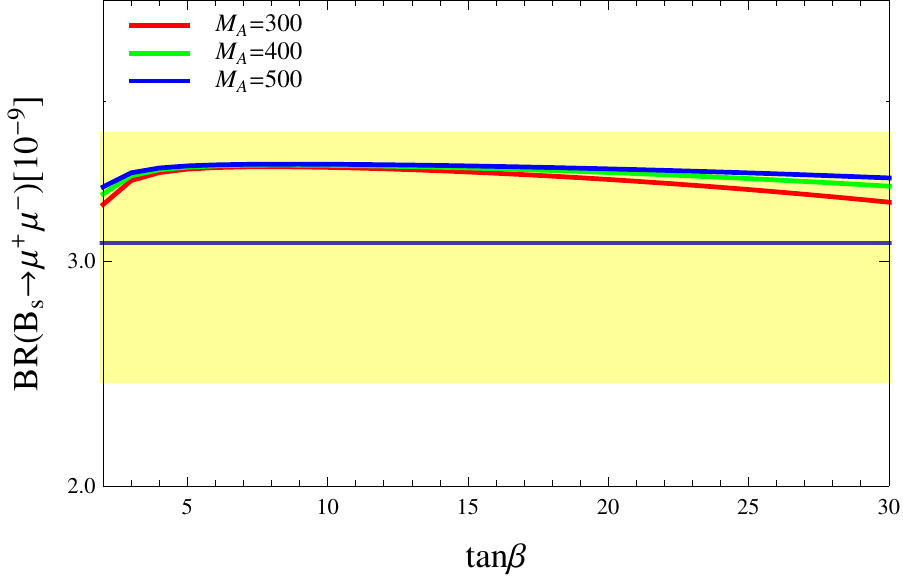}}
\caption{$\mathcal{B}\left(B_{s}\rightarrow\mu^{+}\mu^{-}\right)$ and $\mathcal{B}\left(B\rightarrow X_{s}\gamma\right)$ versus $\tan\beta$ in 2HDM. The $1\sigma$ uncertainty region is shown in yellow. }
\label{2HDM}
\end{figure}

In Fig.\ref{2HDM}, we plot the branching ratios
of $B\rightarrow X_{s}\gamma$ and $B_{s}\rightarrow\mu^{+}\mu^{-}$ with different values of $M_{A}$ in Type II 2HDM. The package {\it SUSY\_Flavor 2.52}\cite{Rosiek:2014sia} is adopted to obtain the numerical results. Notice that in {\it SUSY\_Flavor}, the SM value of $\mathcal{B}(B\rightarrow X_{s}\gamma)$ is evaluated at the NLO to be $0.339\times10^{-3}$, which is about $7.5\%$ larger than the NNLO theoretical prediction of Eq.(\ref{smbsgamma}). Taking this into account, we rescale the experimental bound of $\mathcal{B}(B\rightarrow X_s \gamma)$ to be $(3.69\pm0.24)\times10^{-4}$ in {\it SUSY\_Flavor} in the following analysis. As $\mathcal{B}(B_{s}\rightarrow\mu^{+}\mu^{-})$ is well consistent with the experimental data as shown in Fig.\ref{2HDM}, it implies that in this channel the 2HDM amplitudes with $M_A>300$ GeV are small even in the large $\tan\beta$ region. But $\mathcal{B}\left(B\rightarrow X_{s}\gamma\right)$ is well above the experimental band, which means $M_A\lesssim 400$ GeV
is excluded in Type II 2HDM concerning $B\rightarrow X_{s}\gamma$ decay.

\begin{figure}[h]
\centering
\subfloat[]{\includegraphics[width=6cm]{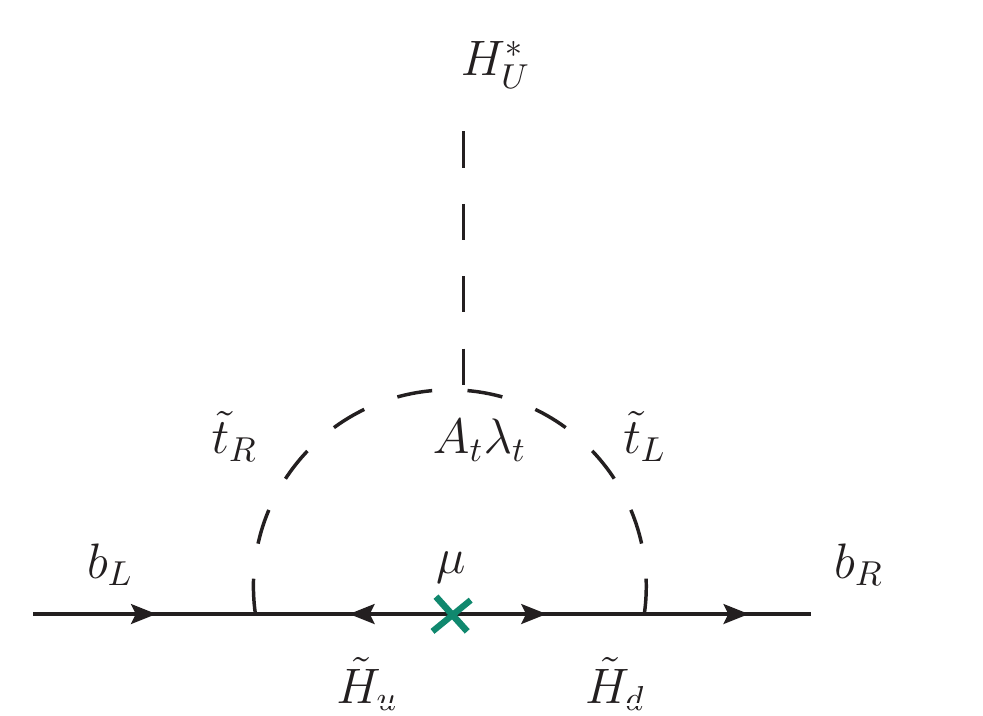}}
\subfloat[]{\includegraphics[width=6cm]{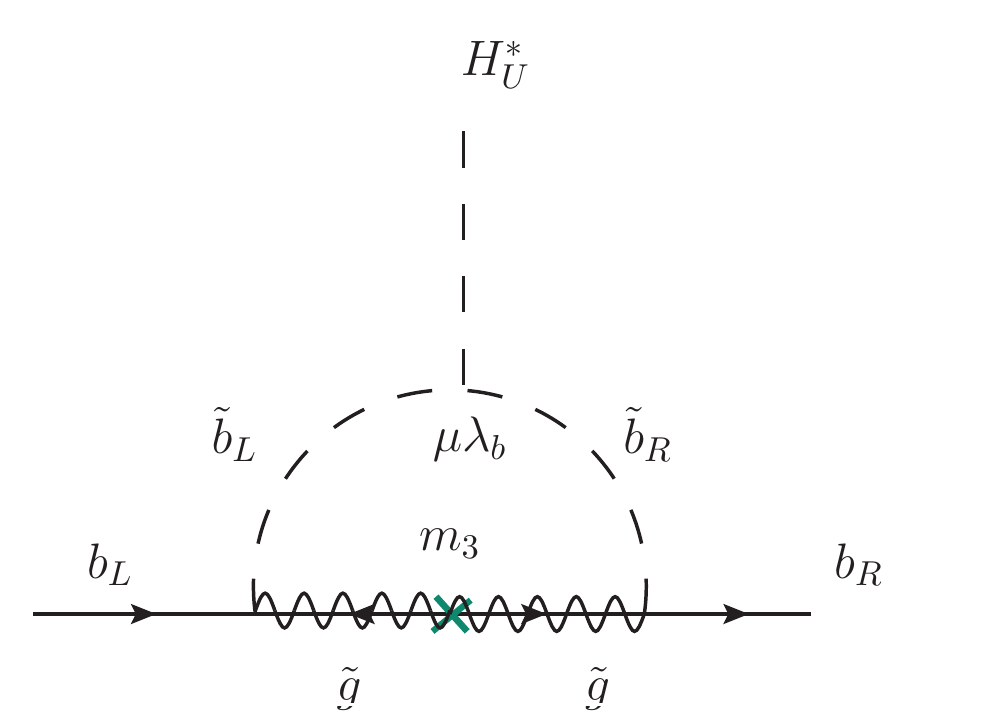}}
\caption{$H_u$ contribution to b-quark mass in MSSM}
\label{bmass}
\end{figure}

MSSM is non-trivial in flavor physics because it contains both
a Type-II 2HDM Higgs sector and sparticles with undetermined masses. For example,
a $b\rightarrow s$ transition could be generated with squark and chargino in the loop. It is well known that
in the large $\tan\beta$ region, SUSY contribution could be significantly enhanced (see, for example, \cite{Babu:1999hn,Choudhury:1998ze,Huang:2000sm}). This enhancement shared by both $B\rightarrow X_{s}\gamma$ and $B{}_{s}\rightarrow\mu^{+}\mu^{-}$ is a result of the mass correction from $Qd^{c}\bar{H}_{u}$ effective vertex shown in Fig.\ref{bmass}.
Therefore we shall focus on the $\tilde{H}\text{-}\tilde{t}$ loop correction in the following.

On the other hand, though a large $b\rightarrow s$ transition is not observed, it does not necessarily mean that
the sparticles should be very heavy. This is because the sparticle contributions might be (partly) canceled by the charged Higgs amplitude.
Additionally, the observed $125$ GeV Higgs could be accounted for in MSSM with light stop masses and large trilinear coupling
for mass splitting. To have a better understanding on the light stop scenario, we simplify
our analysis from the vast MSSM parameter space to a leading $\tilde{\chi}^{\pm}$-$\tilde{q}$
loop and a $\tilde{H}\text{-}\tilde{t}$
loop dominates in this case. The $\tilde{W}$ and
$\tilde{g}$ contributions depend on the undetermined mass insertion $\delta_{IJ}$ in different SUSY breaking patterns. These off-diagonal
terms in the squark mass matrices may lead to FCNC processes and are strongly constrained
in MSSM. So we will not discuss them further in the following.

For $B\rightarrow X_{s}\gamma$ decay, the $\tilde{H}\text{-}\tilde{t}$ loop may interfere either constructively or destructively
with the charged Higgs amplitude depending upon the sign of $\mu A_{t}$\cite{Carena:1994bv,Altmannshofer:2009ne}. As the experimental
result is only slightly larger than the SM prediction, a negative $\mu A_{t}$ is highly preferred if both stop and charged Higgs are relatively light. But for $B{}_{s}\rightarrow\mu^{+}\mu^{-}$ decay, the charged Higgs contribution is small for $M_A>300$~GeV. With negative $\mu A_{t}$, the $\tilde{H}\text{-}\tilde{t}$ loop would interfere constructively with the SM and therefore $\mathcal{B}(B_{s}\rightarrow\mu^{+}\mu^{-})$ is approximately a monotonic increasing function of $\tan\beta$. Actually, BSM contributions which interfere constructively in one process will always interfere destructively in another process, and vice versa.

Let's take a closer look at these decay channels. As a first try, we take $\mu$=$500$~GeV, $\left|A_{t}\right|=2000$~GeV, $m_{t_{L}}$=$2000$~GeV, $m_{t_{R}}$=$500$~GeV and decouple all the other sparticles by assigning very heavy mass. The sign of $A_t$ could be either positive or negative.

\begin{figure}[ht]
\centering
\subfloat[$B\rightarrow X_{s}\gamma$]{\includegraphics[scale=0.8]{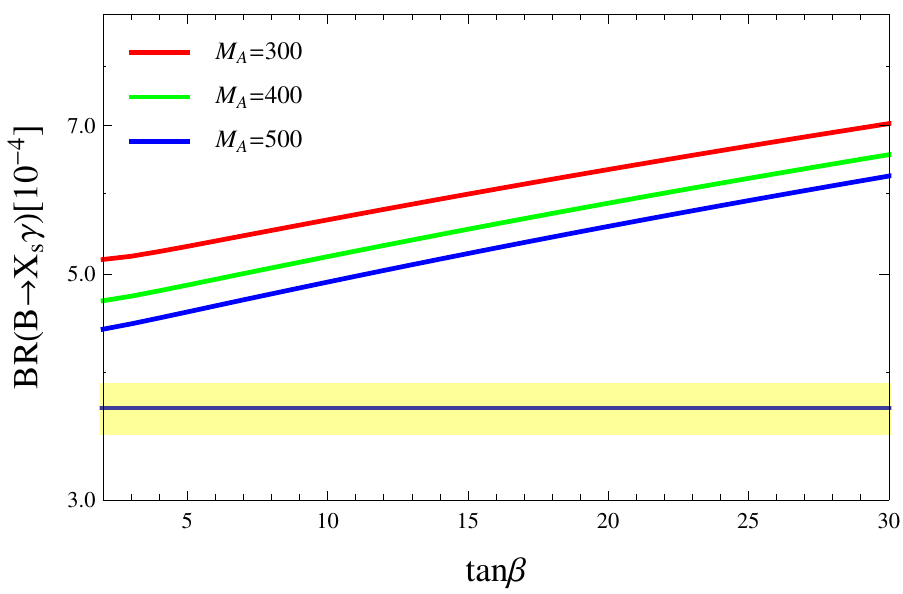}}\quad
\subfloat[$B_{s}\rightarrow\mu^{+}\mu^{-}$]{\includegraphics[scale=0.8]{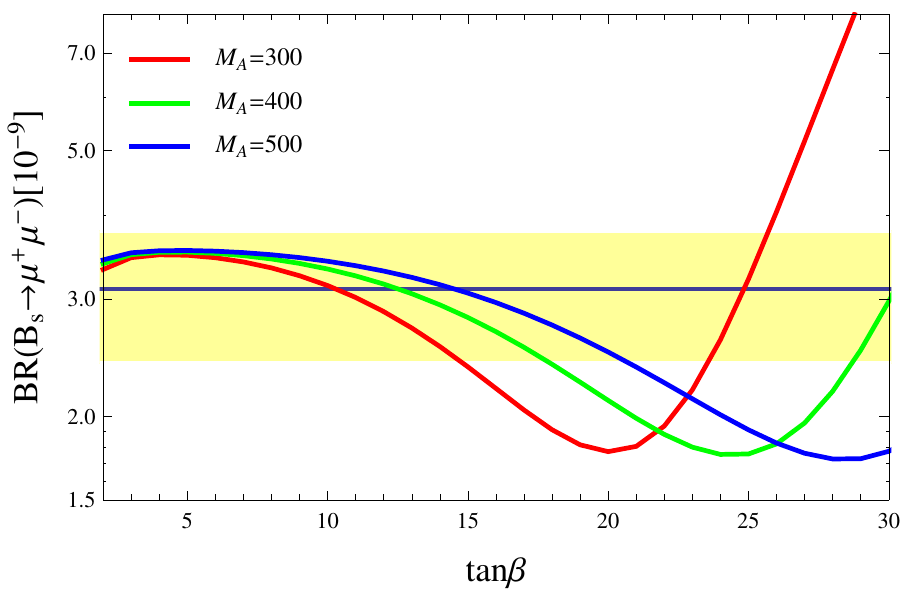}}\\
\subfloat[$B\rightarrow X_{s}\gamma$]{\includegraphics[scale=0.8]{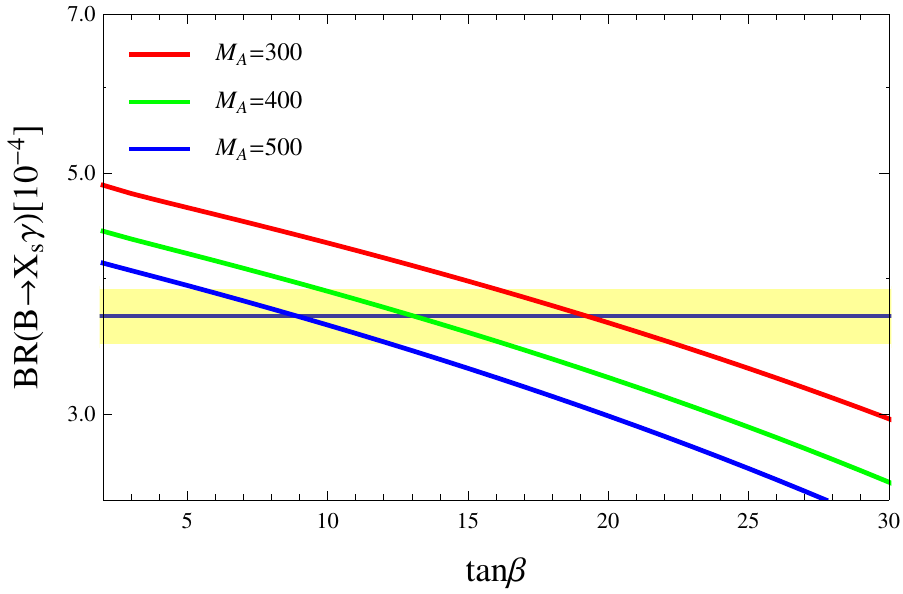}}\quad
\subfloat[$B_{s}\rightarrow\mu^{+}\mu^{-}$]{\includegraphics[scale=0.8]{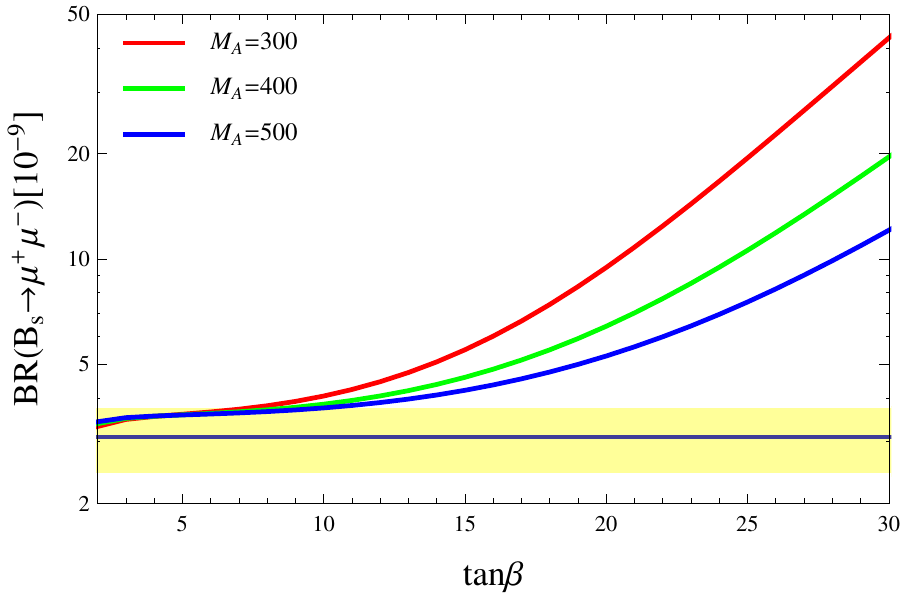}}
\caption{$\mathcal{B}\left(B\rightarrow X_{s}\gamma\right)$ and $\mathcal{B}\left(B_{s}\rightarrow\mu^{+}\mu^{-}\right)$ versus $\tan\beta$ in MSSM with different $A_t$ values. $A_t$ is set to be $2000~\text{GeV}$ in (a)(b) and $-2000~\text{GeV}$ in (c)(d). }
\label{MSSM}
\end{figure}

As shown in Fig.\ref{MSSM}(a), the predicted $B\rightarrow X_{s}\gamma$ decay with positive $A_{t}$ is obviously too large to be consistent with the experimental bound.\footnote{It may be possible to reconcile the theoretical prediction with the experimental data if additional gaugino loop contributions are included with fine-tuned mass and coupling relations.}
For the case of negative $A_{t}$ shown in Fig.\ref{MSSM}(c)(d), these two decay branching ratios could be consistent with the experimental data within $1\sigma$ error separately, but not simultaneously.

In order to find out the viable
parameter space, one may increase the heavy Higgs mass because the beyond standard model
amplitude terms in $B_{s}\rightarrow\mu^{+}\mu^{-}$ are suppressed
by $M_{A}^{2}$.  This will also suppress the contribution of the charged Higgs in $B\rightarrow X_{s}\gamma$
without affecting the negative Higgsino term, which leads to a smaller $\mathcal{B}(B\rightarrow X_{s}\gamma)$.
However, decoupling $H^{0}/H^{\pm}$ somehow goes against our initial purpose to explore a light
extended Higgs sector. Another way is to tune the Higgsino mass parameter $\mu$. The role of
$\mu$ is a little tricky in both decays because it may appear either as a mass insertion in the numerator, or as a propagator in the denominator. If Higgsino is as heavy as multi-TeV,
the $\tilde{H}^{\pm}$-$\tilde{t}$ contribution would clearly be negligible. But
when it is lighter than 1 TeV, there is a region that the branching ratios of both decays can be lowered.
For this reason, a light Higgsino or more accurately a small Higgsino-stop mass ratio can ease the tension and allow a looser $M_{A}$ lower bound.

We then scan $\mu$ parameter to find phenomenologically
preferred region. The scan range is $200~{\rm GeV}\leq\mu\leq2500~{\rm GeV}$
and the sfermion sector is unchanged with a negative $A_{t}=-2500$~GeV. The
heavy Higgs mass $M_{A}$ is set to $600~{\rm GeV}$ and curves with respect to different $\tan\beta$
values are presented.
\begin{figure}[ht]
\centering
\subfloat[$B\rightarrow X_{s}\gamma$]{\includegraphics[scale=0.85]{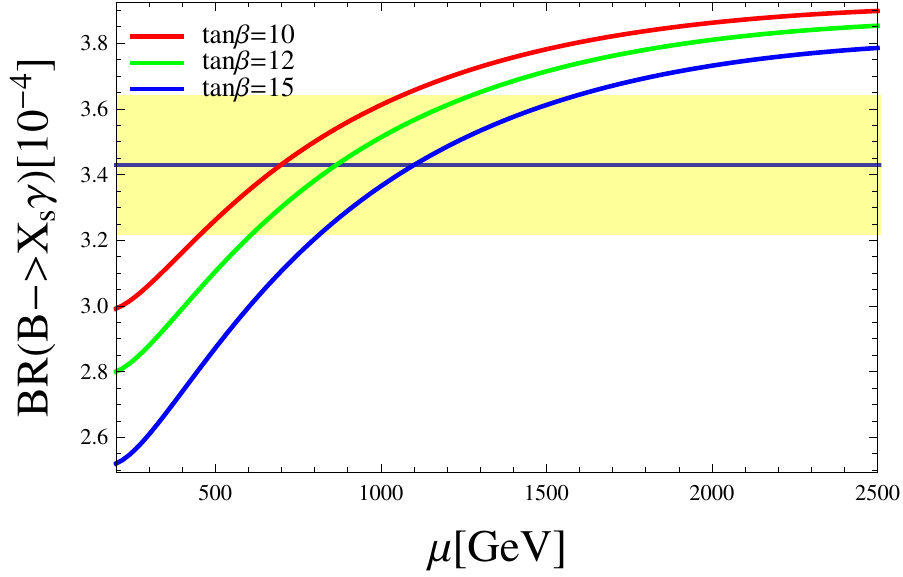}}\quad
\subfloat[$B_{s}\rightarrow\mu^{+}\mu^{-}$]{\includegraphics[scale=0.85]{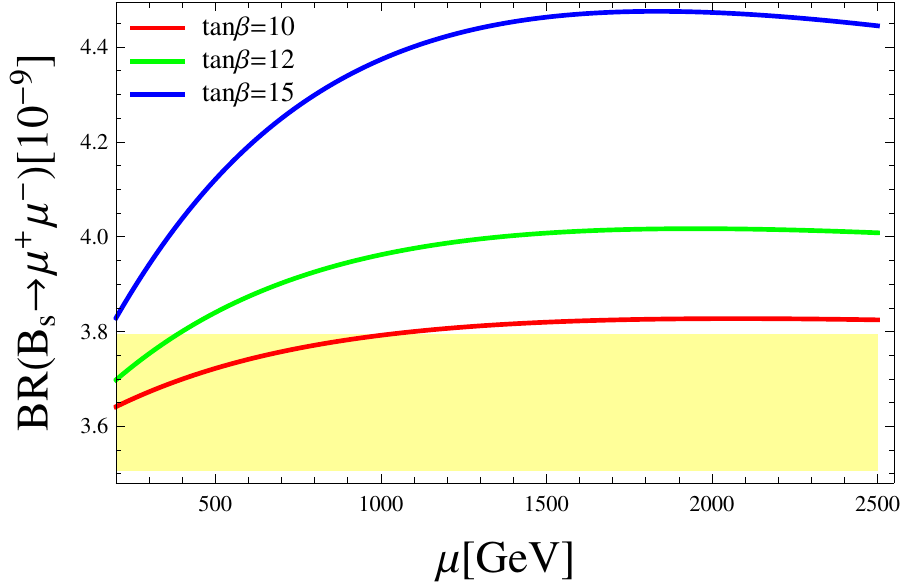}}
\caption{$\mathcal{B}\left(B\rightarrow X_{s}\gamma\right)$ and $\mathcal{B}\left(B_{s}\rightarrow \mu^{+}\mu^{-}\right)$ versus $\mu$. We set $M_A=600~\text{GeV}, m_{\tilde{t}_L}=2000~\text{GeV},$ $m_{\tilde{t}_R}=500~\text{GeV}, A_t=-2500~\text{GeV}$.}
\label{mu}
\end{figure}
Fig.\ref{mu}(a) shows that $\mathcal{B}\left(B\rightarrow X_{s}\gamma\right)$
appears as a monotonic increasing function of $\mu$. This is because the effect of mass insertion $\mu$ is negligible in this decay channel.
When $\mu$ is as large as multi TeV, the branching ratio is almost identical to the 2HDM result as $\tilde{H}^{\pm}$ is
effectively decoupled. The case of $B_{s}\to \mu^+ \mu^-$ decay is somewhat different: we notice a maximum peak with
$\mu$ above $1~{\rm TeV}$ in Fig.\ref{mu}(b), due to the effect of mass insertion $\mu$. This feature is less discussed in previous studies.
Therefore it is possible to choose a small $\mu$ parameter
to simultaneously suppress $\mathcal{B}\left(B\rightarrow X_{s}\gamma\right)$
and $\mathcal{B}\left(B_{s}\rightarrow\mu^{+}\mu^{-}\right)$
to satisfy the experimental bound.

In the next section, we'll numerically show what a survived region with moderate
heavy Higgs mass and moderate $\tan\beta$  is like.

\section{$M_{SUSY}$/$M_{\tilde{t}_1}-\mu$ numerical result}

In order to feature MSSM contribution with a $\tilde{H}^{\pm}$-$\tilde{t}$
loop, we choose six variables in this numerical analysis as
$$\tan\beta, M_{A}, m_{\tilde{t}_{R}}, m_{\tilde{Q}_{3}}, \mu, A_{t}$$
and set the other sfermion soft masses to $3~{\rm TeV}$ with vanishing A-terms
in order to decouple the sub-leading contributions. This $m_{\tilde{t_{R}}}\sim m_{\tilde{Q_{3}}}\ll m_{\tilde{q}}$ circumstance will also avoid the super-GIM suppression in MSSM. Note that $m_{\tilde{b}_{L}}$
is degenerate with $m_{\tilde{t}_{L}}$, so there could be a light
$\tilde{b}_{1}$ in our benchmark, but its contribution is negligible.
As for the gauginos, we set $M_{1}=100~{\rm GeV},M_{2}=M_{3}=2~{\rm TeV}$.

In order to illustrate the properties of the survived region under
the SM-like Higgs bound and flavor physics bound, we implement a comprehensive
scan over the parameter region:
$$200~{\rm GeV}\leq m_{\tilde{Q_{3}}}\leq3.5~{\rm TeV}$$
$$200~{\rm GeV}\leq m_{\tilde{t_{R}}}\leq3.5~{\rm TeV}$$
$$200~{\rm GeV}\leq\mu\leq1~{\rm TeV}$$
$$-4~{\rm TeV}\leq A_{t}\leq-1~{\rm TeV}$$
Since cancellation is necessary in our benchmarks with additional
gaugino contributions, we focus on the negative $A_{t}$ case.
The $B_{s}\rightarrow\mu^{+}\mu^{-}$ bound is too strict for a
scan over $\tan\beta$, so we simply choose two representative $\left(M_{A},\tan\beta\right)$
benchmark from the extended Higgs sector to be
$\left(400,10\right)$ and $\left(600,15\right)$. These benchmarks satisfy the decoupling limit
condition given in \cite{Djouadi:2005gj}, we expect the main constraints from the flavor physics bound and Higgs mass requirement.

In addition to $B\rightarrow X_{s}\gamma$ and $B_{s}\rightarrow\mu^{+}\mu^{-}$,
we restrict the benchmark points with following bounds:
$$124~{\rm GeV}\leq m_{h}\leq127~{\rm GeV}$$
$$0.75\leq\frac{\mathcal{B}\left(h^{0}\rightarrow\tau^{+}\tau^{-}\right)}{\mathcal{B}_{SM}\left(h^{0}\rightarrow\tau^{+}\tau^{-}\right)}\leq1.25$$
$$0.8\leq\frac{\mathcal{B}\left(h^{0}\rightarrow b\bar{b}\right)}{\mathcal{B}_{SM}\left(h^{0}\rightarrow b\bar{b}\right)}\leq1.2$$
$$0.65\leq\frac{\mathcal{B}\left(h^{0}\rightarrow WW,ZZ\right)}{\mathcal{B}_{SM}\left(h^{0}\rightarrow WW,ZZ\right)}$$
$$0.92\times10^{-4}\leq\mathcal{B}\left(B\rightarrow\tau\nu_{\tau}\right)\leq1.36\times10^{-4}$$
We use {\it SUSY\_Flavor 2.52} to compute the B-meson decay branchings and {\it FeynHiggs 2.11.2}\cite{Heinemeyer:1998yj,Heinemeyer:1998np,Degrassi:2002fi,Frank:2006yh,Hahn:2013ria}
to compute the Higgs mass and its decay.

\begin{figure}[H]
\centering
\subfloat[$M_A=400~\text{GeV}, \tan\beta=10$]{\includegraphics[scale=0.8]{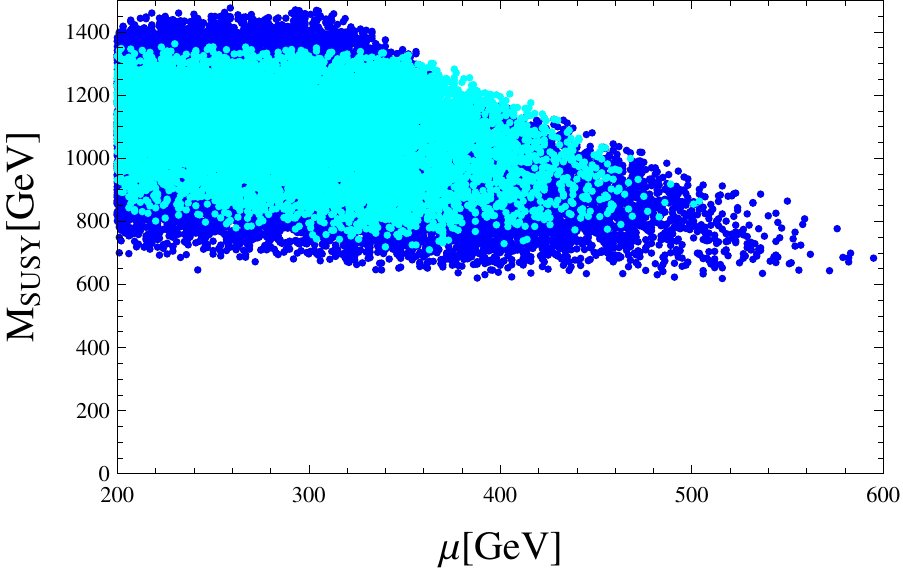}}\quad
\subfloat[$M_A=400~\text{GeV}, \tan\beta=10$]{\includegraphics[scale=0.81]{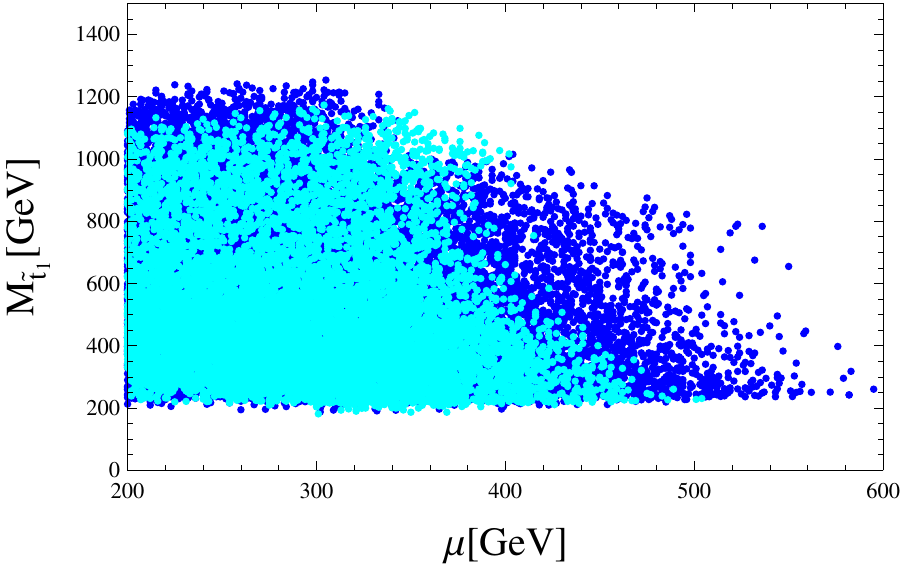}}\\
\subfloat[$M_A=600~\text{GeV}, \tan\beta=15$]{\includegraphics[scale=0.8]{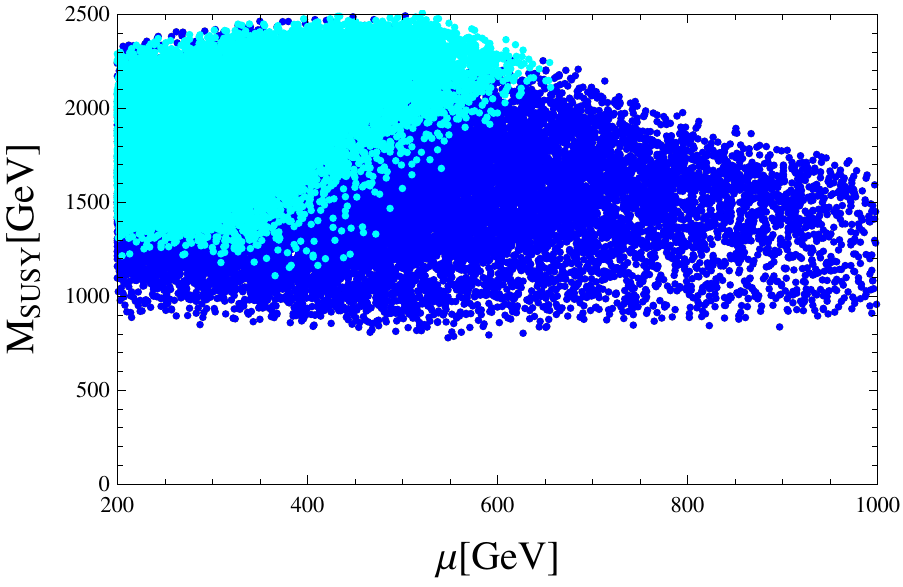}}\quad
\subfloat[$M_A=600~\text{GeV}, \tan\beta=15$]{\includegraphics[scale=0.81]{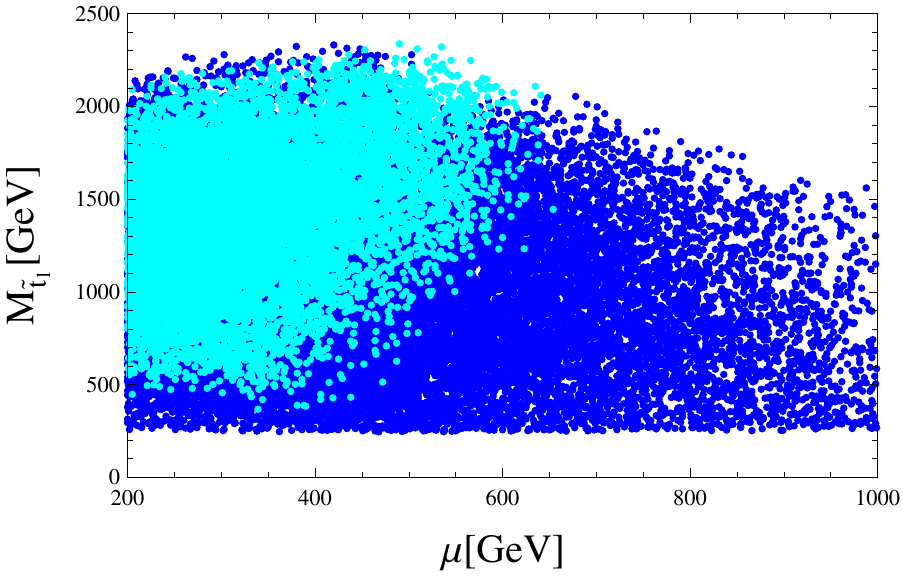}}
\caption{$M_{SUSY}$/$M_{\tilde{t}_1}$ versus $\mu$ is shown with different ($M_A,\tan\beta$) values. The cyan region passes all the bounds while the blue region passes only the $1\sigma$ flavor physics bounds.}
\label{region}
\end{figure}

In Fig.\ref{region}(a)(b), we show the stop mass scale $M_{SUSY}$ and lightest stop mass $M_{\tilde{t}_1}$ versus $\mu$ result when $M_{A}=400~{\rm GeV},\tan\beta=10$. The $m_{h}^{max}$ condition
that $\left|X_{t}\right|=\sqrt{6}M_{SUSY}$ restricts $A_{t}$ within
a range for sufficient $h$ mass correction so that the cyan region which satisfies
all bounds is smaller than the blue region with flavor bound
only. Since we have fixed a $\tan\beta$ and it should locate in the
$B\rightarrow X_{s}\gamma$ safe $\tan\beta$ region, $m_{\tilde{t}_{L,R}}$
are now not arbitrary. The upper bound on $M_{SUSY}$  in Fig.\ref{region}(a) corresponds
to the case when the suppressed $\tilde{H}^{\pm}$-$\tilde{t}$ contribution
fails to cancel $H^{\pm}$-$t$ contribution out. On the
contrary, over cancellation happens if stops are too light. Indeed
we find there's such a lower bound on $M_{SUSY}$ given by $B\rightarrow X_{s}\gamma$
, but the actual boundary in Fig.\ref{region} is from $B_{s}\rightarrow\mu^{+}\mu^{-}$
due to its sensitivity to SUSY contributions. The behaviour with
respect to $\mu$ parameter meets our expectation and the upper limit
is observed just above $450~{\rm GeV}$. In Fig.\ref{region}(b), we find that though $M_{SUSY}$ locates near the  $~{\rm TeV}$ scale, $\tilde{t}_{1}$ and $\tilde{t}_{2}$ could have large mass splitting due to splitted
$\tilde{t}_{L,R}$ mass inputs or large $A_{t}$ in the off-diagonal
mass matrix elements. Thus a light stop is required when the extended Higgs sector does not decouple.

In Fig.\ref{region}(c)(d), we set $\left(M_A,\tan\beta\right)$ larger to $(600,15)$ for a less constrained
region. The discussion above is still valid in this benchmark but
the maximums are larger. More heavy stops survive because one
can always decouple BSM contributions in $B\rightarrow X_{s}\gamma$
by decreasing Higgsino and
charged Higgs contributions together, as long as the
cancellation relation is preserved. Higgsino could be heavier
in this case because larger $M_A$ value works to suppress $\mathcal{B}\left(B_{s}\rightarrow\mu^{+}\mu^{-}\right)$ with its quartic. In the performed numerical analysis, we find such $\mu$ bound is always related to large $\tan\beta$ values.

\begin{figure}[H]
\centering
\subfloat[$M_A=400~\text{GeV}, \tan\beta=20$]{\includegraphics[scale=0.8]{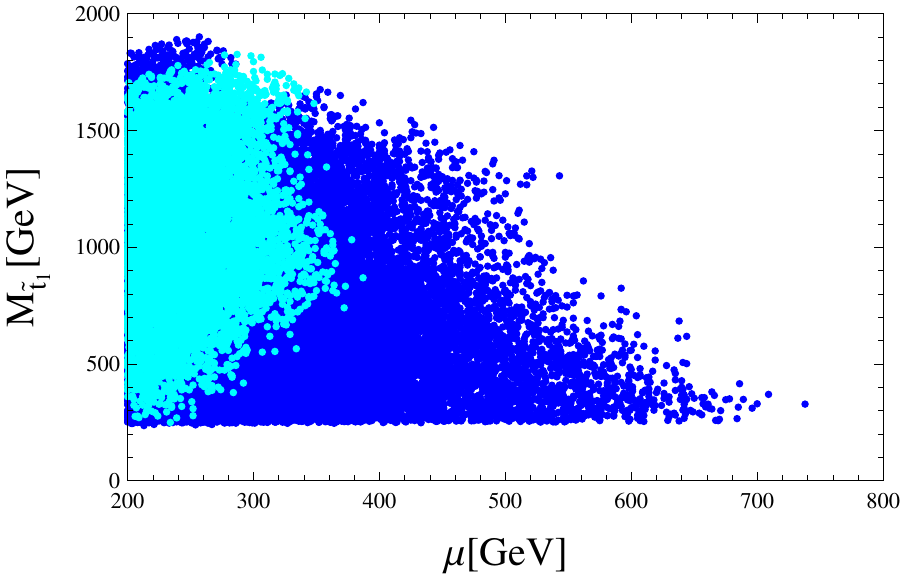}}\qquad
\subfloat[$M_A=600~\text{GeV}, \tan\beta=25$]{\includegraphics[scale=0.815]{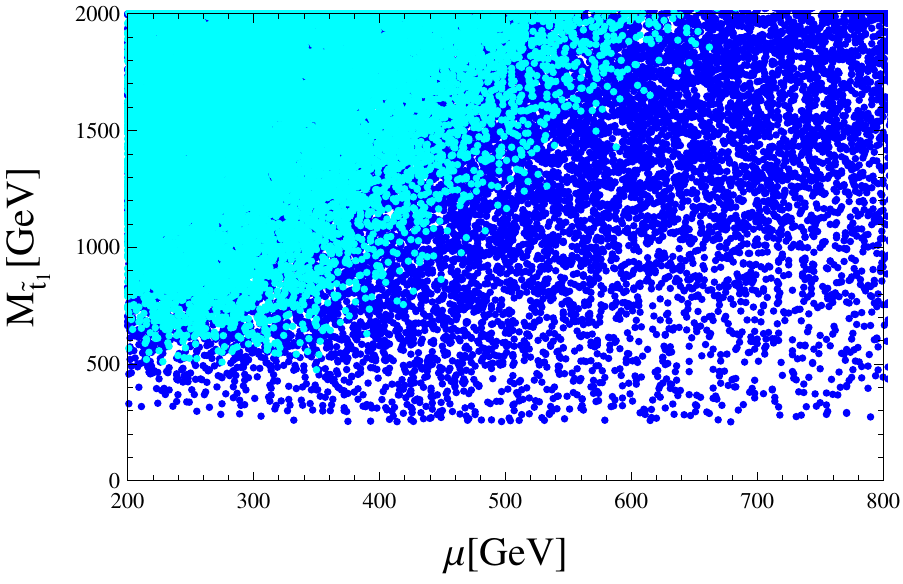}}
\caption{$M_{\tilde{t}_1}$ versus $\mu$ is shown with different ($M_A,\tan\beta$) values. The cyan region passes all the bounds while the blue region passes only the $2\sigma$ flavor physics bounds.}
\label{region_2sigma}
\end{figure}

In Fig.\ref{region_2sigma}, we concisely show the survived region within a $2\sigma$ range. The $\tan\beta$ values are larger than the previous benchmarks in order to present qualitative features of the $\tilde{H}^{\pm}$ and $\tilde{t}_{1}$
mass relation. For a specific $\tilde{t}_1$ mass, there's still a corresponding $\mu$ bound, indicating the tension in $B_{s}\rightarrow\mu^{+}\mu^{-}$ still exists. In the further decoupled case in Fig.\ref{region_2sigma}(b), $M_{A}=600$~GeV, the Higgs contribution to flavor physics is rather small and the flavor constraint can then
be neglected. The only constraint is due to the Higgs mass bound.

We argue that the existence of an upper bound on $\mu$ is an appealing
feature of these benchmarks. Though similar light $\tilde{H}^{\pm}$ requirement occurs in the natural SUSY scenario, which seeks a mild fine-tuning condition, one should note these two confusable results come from totally different motivations. Here we do not emphasise the specific
values of the bound for each benchmarks since this is only a comprehensive
scan and not the full MSSM parameter space has been covered. Detailed
simulation in this light $\tilde{H}^{\pm}$-$\tilde{t}$ scenario is expected with fine-tuned sub-leading contributions.

The light $\tilde{t}_{1}$ in Fig.\ref{region_2sigma}(a) is about $500~{\rm GeV}$ with $\tilde{H}^{\pm}$
mass up to $250~{\rm GeV}$ and that corresponding to the heaviest $\tilde{H}^{\pm}$
is about $1~{\rm TeV}$. Light stops receive stringent constraints from direct search at LHC. For $\tilde{t}_{1}\rightarrow t\tilde{\chi}^{0}_1$, the bound is over $M_{\tilde{t}_1}>600$~GeV. However, if stop decays into Higgsino plus b-jet, the signal encounters large SM background without handle of top reconstruction. If the lightest stop and the charged Higgsino are degenerate at the lower right corner of the cyan regions, the $\tilde{t}_1$ search will be more challenging. Such phenomenology of light $\tilde{t}_{1}$ with Higgsino-like $\tilde{\chi}_{1}^{\pm}$ and its potential to
be found at LHC Run2 is worth further study.

\section{Conclusion}
In this paper, we study a representative region of MSSM parameter space and employ both the flavor physics bound from B-meson rare decays and LHC Higgs constraint. We assume a light extend Higgs initially and then adjust the leading sparticle loop contribution to achieve flavor-safe interference. Strong enhancement caused by $\tan^{n}\beta$ is observed. Rather than suppress the extended Higgs and sparticles contributions or introduce sub-leading terms for cancellation, we find that as long as the sparticle $\tilde{H}^{\pm}$ and $\tilde{t}$ in the loop are light, the assumption of light $H^{0}/H^{\pm}$ is still practicable.

\section{Acknowledgement}
The work is supported in part by the National Science Foundation of China (11135006,  11275168, 11422544, 11075139, 11375151, 11535002) and the Zhejiang University Fundamental Research Funds for the Central Universities. KW is also supported by Zhejiang University K.P.Chao High Technology Development Foundation.
GZ is also supported by the Program for New Century Excellent
Talents in University (Grant No. NCET-12-0480). LZ is grateful to Rijun Huang, Qingjun Jin and Wolfgang Altmannshofer for useful discussions.

\end{document}